\documentclass[12pt]{iopart}
\usepackage[dvips]{graphics,color}
\usepackage{graphicx}
\usepackage{dcolumn}
\usepackage{bm}
\usepackage{epsfig}

\newcommand{\beq}{\begin{equation}}
\newcommand{\eeq}{\end{equation}}
\newcommand{\beqa}{\begin{eqnarray}}
\newcommand{\eeqa}{\end{eqnarray}}
\newcommand{\bd}[1]{ \mbox{\boldmath $#1$}}

\begin{document}
\def\ii{\'\i}

\title{
A proposal of Quantization in flat space-time with a minimal length present
}

\author{Peter O. Hess}

\address{
Instituto de Ciencias Nucleares, UNAM, Circuito Exterior, C.U.,\\
A.P. 70-543, 04510 M\'exico, D.F., Mexico
}
\ead{hess@nucleares.unam.mx}
\begin{abstract}
The 4-dimensional space-time is extended to pseudo-complex coordinates. 
Proposing the standard quantization rules in this extended space, the ones
for the 4-dimensional sub-space acquire, as one solution, the commutation relations with non-commuting
coordinates. This demonstrates that the algebraic extension keeps the
simple structure of Quantum Mechanics, while it also introduces an effective quite
involved structure in the 4-dimensional sub-space. The similarities to H. S. Snyder's work,
a former proposal to include the effects of a minimal length, are exposed.
The first steps to pseudo-complex Quantum Mechanics in 1-dimension are outlined,
awaiting still the interpretation of some new emerging structures.
As an example, two waves, out of phase by
90 degrees, are added which classically annihilate each other, while
in the pseudo-complex description there is a non-zero amplitude. 

\end{abstract}

\maketitle

\section{Introduction}

There is a long tradition to investigate the effects of a minimal length in space-time,
leading to non-commutative behavior of the coordinates. The first, to my knowledge, who studied
the effects of a minimal length, was H. S. Snyder \cite{Snyder}.
The non-commutativity of coordinates is often imposed defining the 
{\it Moyal product} \cite{Moyal}. A complete reference list would be too long, thus, I
will mention only some of them, relevant in this contribution (for more, please consult the
reference list of the cited publications).
More recent applications can be found in \cite{Moffat,Nicolini,obregon1,obregon2}. This Moyal product introduces a complicated
structure, which render its application difficult. To mention is also the 
{\it Theory of Quantum Loops} \cite{smolin} and works related to it \cite{amelini}, which 
predict different velocities of light at different frequencies, due to the presence of
a minimal length \cite{amelini}. 
{\it String Theory} also contains a minimal length \cite{string}. Though, this theory 
is believed to describe the physics in the universe, 
its formulation is quite complicated and it is useful to explore other {\it effective descriptions}.
For an overview on {\it String Theory} and {\it Quantum Loops}, on a popular level and with more recent references on both topics, 
see \cite{smolin-book}.
All these theories have an involved structure and 
include the violation of the Lorentz symmetry. 
In easy terms, the last point is mainly understood as follows:
When a minimal length is present in nature, also called {\it granulation of space-time},
it should be a natural constant. However, a pure Lorentz transformation changes the length
element, not leaving it invariant, i.e., Lorentz symmetry
has to be violated.

This assumes a physical length! However, when this minimal length is introduced as a parameter
into the theory, it is not affected by a Lorentz transformation! This was pointed out
in \cite{schuller,pcFT,pcGR1} and did lead to a simplified structure of field theories and
General Relativity. Also in other contributions 
\cite{caianiello,brandt1,beil1} 
the minimal length is introduced in this manner.
For example, in \cite{schuller} and \cite{pcFT} it is shown that
there is a regularized version of field theories according to Pauli-Villars, but
gauge symmetry is still preserved! In \cite{pcGR1} a pseudo-complex version of General
Relativity was proposed, which leads again to a simplified structure of General Relativity
with an additional contribution due to which the event horizon of a former black hole
disappears. In conclusion, one can introduce a minimal length but still
keep a continuous space-time.

In this contribution I want to demonstrate that extending algebraically the space-time
coordinates to {\it pseudo-complex} (pc) variables, allows to follow 
the standard quantization procedure in a higher 8-dimensional space, while
in the 4-dimensional sub-space the coordinates are non-commutative. 
The mathematics is not more complicated than for complex variables and functions.
This avoids 
the use of involved structures as the Moyal product, quantum loops or string theory 
and keeps physics simple.
We show that the proposed quantization procedure is similar to the one in \cite{Snyder}, though 
differences appear also. What will be presented here is an effective theory, which probably
is a limit of the {\it Theory of Quantum Loops} and/or {\it String Theory}.

The paper is organized as follows: In section 2, I will resume the main mathematical
properties of pc-variables and functions. Also the structure of the Lorentz group
and its algebra is discussed,
showing the conservation of this symmetry in the higher dimensional space. 
In section 3, some basic properties of the geometry in flat space are resumed and the interpretation of the pseudo-imaginary  component of the coordinate is given.
In section 4,
the quantization proposal will be
presented and shown that though a minimal length appears in the theory its description
is still very simple.
In section 5 we compare it to Snyder's work in \cite{Snyder}. In section 6 
the first steps of a 1-dimensional pseudo-complex Quantum Mechanics are presented. It
serves to outline the main ideas but will not be complete due to some
open questions in the interpretation of emerging new structures.
A simple
example will be presented, namely the sum of two plane waves out of phase by $\pi$ in the
{\it x-space} (also denoted as the {\it 4-dimensional subspace}). Classically the amplitude of the
sum of the two waves vanish.
It is shown that this sum does produce a non-zero amplitude,
though still too small to be measured yet. The main motivation is to point out 
that a Quantum Mechanics in the extended pc-space retains a simple structure
and that there are differences, which {\it in principle} can be measured. 
An algebraic extension, which leads to a simple structure, may enlighten possible ways to quantize 
with a minimal length present and may show some possibilities on how to quantize
curved space-time.
In section 7 conclusions are drawn. In the appendix, I
deduce the commutation relations of the Lorentz operators in the 4-dimensional sub-space,
showing that the Lorentz symmetry is effectively broken there.

\section{Pseudo-complex variables, pseudo-complex functions and Lorentz symmetry}

The notation in this section is adapted to the use of coordinates in four dimensions.

The pc-coordinates are defined as

\beqa
X^\mu & = & x^\mu + I y^\mu
\nonumber \\
{\rm with}~I^2 & = & 1
~~~.
\label{pc-1}
\eeqa
This explains the name {\it pseudo}, as introduced in \cite{schuller}. There are, however, 
other names in literature \cite{pcGR1} as {\it hyperbolic, hyper-complex, para-complex and more}.

An alternative representation for $X^\mu$ is found, introducing

\beqa
\sigma_\pm & = & \frac{1}{2}\left( 1 \pm I \right)
\nonumber \\
\sigma_\pm^2 & = & \sigma_\pm ~,~ \sigma_+\sigma_- ~=~ 0
~~~.
\label{pc-2}
\eeqa
With (\ref{pc-2}), the coordinates acquire the form

\beqa
X^\mu & = & X_+^\mu \sigma_+ + X_-^\mu \sigma_+
\nonumber \\
X_\pm^\mu & = & x^\mu \pm y^\mu
~~~.
\label{pc-3}
\eeqa

Due to second relation in (\ref{pc-2}), $\lambda \sigma_+$ multiplied with $\mu \sigma_-$ gives zero, where
$\lambda$ and $\mu$ are arbitrary. This is the standard 
definition for the existence of
a {\it zero-divisor}. Thus, the pc-variables do not form a field but rather a ring.

A further important consequence is that mathematical manipulations within the 
{\it zero-divisor} components can be applied independently. For example

\beqa
X^\mu X^\nu & = & \left( X_+^\mu \sigma_+ + X_-^\mu \sigma_-\right)
\left( X_+^\nu \sigma_+ + X_-^\nu \sigma_-\right)
\nonumber \\
& = & X_+^\mu X_+^\nu \sigma_+ + X_-^\mu X_-^\nu \sigma_-
\label{pc-4}
\eeqa
and similar for any product of functions

\beqa
F(X)G(X) & = & F_+(X_+)G_+(X_+)\sigma_+ + F_-(X_-)G_-(X_-) \sigma_-
~~~.
\label{pc-5}
\eeqa
The complex conjugate of a pc-variable is defined as

\beqa
X^{\mu~*} & = & x^\mu - I y^\mu ~=~ X_+^\mu \sigma_- + X_-^\mu \sigma_+
~~~.
\label{pc-6}
\eeqa
Using $\sigma_\pm^* = \sigma_\mp$,
the norm is

\beqa
\mid X^\mu \mid & = & X^{\mu~*} X^\mu ~=~ \left( x^\mu\right)^2 - \left( y^\mu\right)^2
~=~X_+^\mu X_-^\mu
~~~.
\label{pc-7}
\eeqa
Elements in the zero-divisor satisfy $y^\mu = \pm x^\mu$, or equivalently
$X_+^\mu $ or $X_-^\mu$ equal to zero.

A different definition for the norm is the {\it euclidean norm}

\beqa
\mid\mid X^\mu \mid\mid_E & = & \sqrt{
\mid X^\mu_R\mid^2+\mid X^\mu_I\mid^2}
~~~,
\label{pc-7-a}
\eeqa
with $X^\mu_R=x^\mu$ as the pseudo-real and $X^\mu_I=y^\mu$ the pseudo-imaginary
component of $X^\mu$. This definition will play an important role
later on, defining probabilities in 1-dimensional Quantum Mechanics.

Also differentiation and integrals can be defined in complete
analogous manner as in standard complex analysis, as long as the line of
the zero-divisor is not crossed. 
For details, please consult \cite{schuller,pcFT,pcGR1,anto}.

In what follows, the algebra of the Lorentz transformation in pc-space are discussed:
The generators of the pc-Lorentz group are given by
the antisymmetric operators

\beqa
L_{\mu\nu} & = & X_\mu P_\nu - X_\nu P_\mu
\nonumber \\
& = & 
\left( X^+_\mu P^+_\nu - X^+_\nu P^+_\mu \right) \sigma_+
+ \left( X^-_\mu P^-_\nu - X^-_\nu P^-_\mu \right) \sigma_-
\nonumber \\
& = & L^+_{\mu\nu} \sigma_+ + L^-_{\mu\nu} \sigma_- 
~~~,
\label{pc-lorentz-generators}
\eeqa
where $X_\mu$ and $P_\nu$ are the pc-coordinates and -momenta, respectively.
The division into the zero-divisor components is also given
and $L^\pm_{\mu\nu}$ are the corresponding generators of
the Lorentz transformation in each sector. By construction
$L_{\mu\nu}^+$ and $L_{\mu\nu}^-$ commute.

A finite pc-Lorentz transformation has the form

\beqa
e^{-i\omega^{\mu\nu}L_{\mu\nu}} & = &
e^{-i\omega_+^{\mu\nu}L^+_{\mu\nu}}\sigma_+ 
+ e^{-i\omega_-^{\mu\nu}L^-_{\mu\nu}} \sigma_-
~~~,
\label{finite-pc-trafo}
\eeqa
having in each zero divisor component a Lorentz transformation and
thus a Lorentz group $SO_\pm (3,1)$.
The parameters of the transformation are also pseudo-complex and have the structure

\beqa
\omega^{\mu\nu} & = & \omega_+^{\mu\nu} \sigma_+ + \omega_-^{\mu\nu} \sigma_-
~=~ \omega_R^{\mu\nu} + I \omega_I^{\mu\nu}
~~~.
\label{omega}
\eeqa

Because the expression in the $\sigma_+$ sector commutes with the one in the
$\sigma_-$ sector (remember that $\sigma_+\sigma_- = 0$), the group structure is given by

\beqa
SO_{pc}(3,1) & = & SO_+(3,1) \otimes SO_-(3,1)
~~~.
\label{group-structure}
\eeqa
The pc-Lorentz group is therefore a direct product of
two independent Lorentz groups. 

The algebra of the pc-Lorentz group is directly obtained and the result is

\beqa
\left[ L_{\mu\nu}, L_{\lambda\delta}\right] & = &
\left[ X_\mu P_\nu - X_\nu P_\mu , X_\lambda P_\delta - X_\delta P_\lambda \right]
\nonumber \\
& = & i\hbar\left[
\eta_{\nu\lambda}L_{\delta\mu} 
+\eta_{\nu\delta}L_{\mu\lambda}
+\eta_{\mu\lambda}L_{\nu\delta}
+ \eta_{\mu\delta}L_{\lambda\nu}
\right]
\label{sol-12}
\eeqa
and the same for each zero-divisor component, i.e., $L_{\mu\nu}$
substituted by $L^\pm_{\mu\nu}$. The $\eta_{\mu\nu}$ is the metric in Minkowsky space,
with the signature $(+---)$.

{\it The Lorentz symmetry is thus maintained on the level of the 8-dimensional
space!}

In \cite{schuller} a pc-field theory was proposed, which keeps the linear
structure of the field theory but has regularized propagators a la Pauli-Villars. 
In \cite{pcFT}
the pc-field theory was explored extensively and shown that also the gauge
symmetries are conserved. 

Thus, it was shown again that a pc-description leads
to an easier treatment of seemingly complicated structures.

\section{Flat space-time and the interpretation of the pseudo-complex components}

The metric in flat space-time is used, as introduced above.
Therefore, $g^\pm_{\mu\nu}$ is given by

\beqa
g^\pm_{\mu\nu} & = & \eta_{\mu\nu}
~~~,
\label{eq-1}
\eeqa
which implies that also the pc-metric is real, i.e. the same in both zero-divisor components:

\beqa
g_{\mu\nu}(X) & = & g^+_{\mu\nu} (X_+) \sigma_+ + g^-_{\mu\nu} (X_-) \sigma_-
\nonumber \\
& = & \eta_{\mu\nu} \left(\sigma_+ + \sigma_- \right) ~=~ \eta_{\mu\nu}
~~~,
\label{eq-1a}
\eeqa
having used $\left( \sigma_+ + \sigma_- \right)=1$

It follows that
upper and lower indices can be lowered and raised  via

\beqa
X_\mu & = & \eta_{\mu\nu}X^\nu
\nonumber \\
X^\mu & = & \eta^{\mu\nu}X_\nu
\label{eq-2}
\eeqa
and the same for $X^\pm_\mu$ and $X_\pm^\mu$.

For the pseudo-complex length element we obtain, with $dX^\mu = dx^\nu +I dy^\mu $,

\beqa
d\omega^2 & = & \eta_{\mu\nu} dX^\mu dX^\nu
\nonumber \\
& = & \eta_{\mu\nu}
\left( dx^\mu dx^\nu + dy^\mu dy^\nu \right) + 2I 
\eta_{\mu\nu} dx^\mu dy^\nu 
~~~,
\label{eq-3}
\eeqa
where the symmetry $\eta_{\mu\nu}=\eta_{\nu\mu}$ was exploited.

The physical requirement is that a {\it length element should be real}.
Thus, the pseudo-imaginary term in (\ref{eq-3}) has to vanish:

\beqa
\eta_{\mu\nu} dx^\mu dy^\nu & = & dx^\mu dy_\mu ~=~ 0
~~~.
\label{eq-4}
\eeqa
which is solved by

\beqa
y^\mu & \sim u^\mu ~=~ \frac{dx^\mu}{ds}
~~~,
\label{eq-5}
\eeqa
where $u^\mu$ are the components of the 4-velocity
and $ds$ is the length element which might be the eigentime. Note, that (\ref{eq-4}) 
with (\ref{eq-5}), give the dispersion relation, which is thus a consequence
of the reality condition of $d\omega^2$. 

Due to dimensional reasons ($y^\mu$ must have units of length), we finally get the identification

\beqa
y^\mu & = & \frac{l}{c} u^\mu
~~~,
\label{eq-6}
\eeqa
where $l$ is the minimal length. Thus, in flat space there is no problem to identify
the pc-component of $X^\mu$ with the 4-velocity. (In a curved space this is a problem because
$x^\mu$ and $u^\mu$ transform differently, the first variable being in a manifold and the
second one in its tangent space.) Furthermore, due to (\ref{eq-6}) a {\it minimal length
parameter} is introduced into the theory

For a while, this will be the last time to express the pseudo-imaginary component of the coordinate
in terms of the 4-velocity. The above considerations served to show that $y^\mu$ has to have
a small value, proportional to $l$. 
This gives us the reason to rename the pseudo-imaginary component as

\beqa
y^\mu & \rightarrow & \left( \frac{l}{c}\right) y ^\mu
~~~,
\label{eq-6-b}
\eeqa
where the new $y^\mu$ is the $u^\mu$ above.

We also require that this result is symmetric under the canonical
transformation which interchanges the coordinates and momenta.
This is similar to {\it Born's proposed reciprocity} \cite{born1,born2},
namely that the coordinates and momenta should appear with equal rights and
the equations have to show a symmetry, interchanging the coordinates with the momenta.

Extracting the scale factor $\left( \frac{l}{c} \right)$,
this implies

\beqa
X^\mu & = & x^\mu + I \frac{l}{c}y^\mu
\nonumber \\
P_\mu & = & p^x_\mu + I \frac{l}{c}p^y_\mu
~~~.
\label{eq-6-1}
\eeqa

\section{A proposal for Quantization}

The proposed rules of quantization are defined in the pc-space, namely

\beqa
\left[ X^\mu , P_\nu \right] & = & i\hbar \delta_{\mu\nu}
\nonumber \\
\left[ X^\mu , X^\nu \right] & = & 0
\nonumber \\
\left[ P_\mu , P_\nu \right] & = & 0
~~~.
\label{eq-7}
\eeqa

These quantization rules, e.g. for the spatial components $X^\mu$, look alike
to the standard ones, however now in the pc-space.
Due to the mathematical properties of pc-variables, the same holds for {\it each
component} in $\sigma_+$ and $\sigma_-$, just by adding in (\ref{eq-7}) the
indices $\pm$ to the coordinates and momenta. 

(\ref{eq-7}) justifies to identify the momentum operator $P_\nu$ with
$\frac{\hbar}{i}\frac{\partial}{\partial X^\nu}$, where a pc-derivative is meant
\cite{anto}, which is applied in the
same way as for real variables. For the zero-divisor components this implies
$P^\pm_\nu$ = $\frac{\hbar}{i}\frac{\partial}{\partial X_\pm^\nu}$. The identification with a derivative
with respect to a coordinate will not be true any more when we restrict to the x-space,
because the commutation relations with the momentum operators will be in general different.

In what follows, the commutation relations of $x^\mu$, $y^\mu$, $p_\mu^x$ and $p_\mu^y$
are deduced:

\vskip 0.5cm
\noindent
(1) $\left[ X^\mu , X^\nu \right] = 0$:

This is equivalent to

\beqa
0 ~=~ \left[ X_+^\mu \sigma_+ + X_-^\mu \sigma_-,X_+^\nu \sigma_+ + X_-^\nu \sigma_- \right]
& = &
\left[ X_+^\mu , X_+^\nu \right] \sigma_+ +  \left[ X_-^\mu , X_-^\nu \right] \sigma_-
~~~,
\nonumber \\
\label{eq-9}
\eeqa
from which follows

\beqa
\left[ X_\pm^\mu , X_\pm^\nu \right] & = & 0
~~~.
\eeqa

In a similar fashion, the other commutation rules result, i.e.,

\beqa
\left[ P^\pm_\mu , P^\pm_\nu \right] & = & 0
\nonumber \\
\left[X_\pm^\mu , P^\pm_\nu \right] & = & i\hbar \delta_{\mu\nu}
~~~.
\label{eq-10}
\eeqa

The $X_\pm^\mu$ are related by definition to the $x^\mu$ and $y^\mu$ via

\beqa
X_\pm^\mu & = & x^\mu \pm \frac{l}{c}y^\mu
~~~.
\label{eq-11}
\eeqa
This gives

\beqa
\left[ x^\mu \pm \frac{l}{c}y^\mu , x^\nu \pm \frac{l}{c}y^\nu \right] & = &
\left(
\left[ x^\mu , x^\nu \right] + \left(\frac{l}{c}\right)^2\left[ y^\mu , y^\nu \right] \right)
\pm \left(\frac{l}{c}\right)\left( \left[ x^\mu , y^\nu \right] + \left[ y^\mu , x^\nu \right] \right)
\nonumber \\
& = & 0
~~~.
\label{eq-12}
\eeqa
Taking the sum and the difference of these two equations (for the positive and negative sign),
we obtain

\beqa
\left[ x^\mu , x^\nu \right] + \left(\frac{l}{c}\right)^2\left[ y^\mu , y^\nu \right] & = & 0
\nonumber \\
\left[ x^\mu , y^\nu \right] + \left[ y^\mu , x^\nu \right] & = & 0
~~~.
\label{eq-13}
\eeqa

Analogous we proceed for the other relations.

\vskip 0.5cm
\noindent
(2) $\left[ P^\pm_\mu , P^\pm_\nu \right] = 0$:

Here, we have only to substitute in the former calculation $X_\pm^\mu$ by $P^\pm_\mu$,
yielding
\beqa
\left[ p^x_\mu , p^x_\nu \right] + 
\left(\frac{l}{c}\right)^2\left[ p^y_\mu , p^y_\nu \right] & = & 0
\nonumber \\
\left[ p^x_\mu , p^y_\nu \right] + \left[ p^y_\mu , p^x_\nu \right] & = & 0
~~~.
\label{eq-14}
\eeqa

As the last relation we investigate

\vskip 0.5cm
\noindent
(3) $ \left[ X_\pm^\mu , P^\pm_\nu \right] = i\hbar \delta_{\mu\nu}$:

We have

\beqa
&
\left[ x^\mu \pm \frac{l}{c}y^\mu , p^x_\nu \pm \frac{l}{c}p^y_\nu \right]
&
\nonumber \\
&
\left(
\left[ x^\mu ,p^x_\nu \right]
+
\left(\frac{l}{c}\right)^2\left[ y^\mu ,p^y_\nu \right]
\right)
\pm \left(\frac{l}{c}\right)
\left(
\left[ x^\mu ,p^y_\nu \right]
+
\left[ y^\mu ,p^x_\nu \right]
\right)
&
\nonumber \\
&
= i\hbar \delta_{\mu\nu}
~~~.
\label{eq-16}
\eeqa
Taking again the sum and the difference of both equations (of the positive and negative
sign), gives

\beqa
\left[ x^\mu , p^x_\nu \right] + 
\left(\frac{l}{c}\right)^2\left[ y^\mu , p^y_\nu \right]
& = & i\hbar \delta_{\mu\nu}
\nonumber \\
\left[ x^\mu , p^y_\nu \right] + \left[ y^\mu , p^x_\nu \right]
& = & 0
~~~.
\label{eq-17}
\eeqa

\vskip 1cm
To resume, we have the following quantization relations for the pseudo-real and
pseudo-imaginary components of the coordinates and momenta:

\beqa
\left[x^\mu , x^\nu \right] & = & 
-\left(\frac{l}{c}\right)^2\left[ y^\mu , y^\nu \right]
\nonumber \\
\left[ x^\mu , y^\nu \right] & = & - \left[y^\mu , x^\nu \right]
\nonumber \\
\left[p^x_\mu , p^x_\nu \right]& = & 
-\left(\frac{l}{c}\right)^2
\left[ p^y_\mu , p^y_\nu \right]
\nonumber \\
\left[ p^x_\mu , p^y_\nu \right] & = & - \left[p^y_\mu , p^x_\nu \right]
\nonumber \\
\left[x^\mu , p^x_\nu \right]
& = & i\hbar \delta_{\mu\nu}
 - 
\left(\frac{l}{c}\right)^2\left[y^\mu , p^y_\nu \right]
\nonumber \\
\left[ x^\mu , p^y_\nu \right] & = & - \left[y^\mu , p^x_\nu \right]
~~~.
\label{eq-18}
\eeqa

The pc-description guarantees that in the 8-dimensional pc-space the
Lorentz symmetry is maintained. However, the situation {\it apparently changes}
when the generators of the Lorentz group are proposed to be \cite{Snyder}

\beqa
{\cal L}_{\mu\nu} & = & x_\mu p^x_\nu -x_\nu p^x_\mu
~~~.
\label{L1}
\eeqa
As shown in APPENDIX A, the commutator of these operators are of the form

\beqa
\left[ {\cal L}_{\mu\nu},{\cal L}_{\lambda\delta}\right]
& = & 
i\hbar \left( \eta_{\nu\lambda} {\cal L}_{\delta\mu}
+\eta_{\delta\nu} {\cal L}_{\mu\lambda}
+\eta_{\mu\lambda} {\cal L}_{\nu\delta}
+\eta_{\mu\delta} {\cal L}_{\lambda\nu} \right)
\nonumber \\
&& + \left(\frac{l}{c}\right)^2 {\bd F}_{\mu\nu\lambda\delta}\left( x,p^x,y,p^y\right)
~~~,
\label{L2}
\eeqa
where ${\bd F}$ is an operator function in the coordinates and momenta.

The first contribution is what one expects 
for the generators of the Lorentz group.
and the second one (second line)
is of the order of $\left(\frac{l}{c}\right)^2$, i.e., {\it a possibly very small number},
if it is for example of the order of the Planck length ($10^{-33}$cm). Nevertheless, when $l$ is much larger than that, one can hope to see possible effects.

The result implies that the Lorentz symmetry {\it appears} to be violated, as one
normally assumes when a minimal length is  introduced. 
Remember however, the Lorentz symmetry
is maintained in the 8-dimensional space!

{\it A general rule follows, namely that 
all calculations are performed in the 8-dimensional space and only at the end the 
real valued observables are extracted.}

This result shows again, that extending the four dimensional space-time to an eight
dimensional pc-space-time conserves the simple structure of Quantum Mechanics,
while in the four dimensional real space-time more
complicated expressions are obtained due to the appearance of the minimal length.

\section{Consequences and relation to H. S. Snyder's proposition}

In this section we investigate 
consequences of (\ref{eq-18}) and its relation to the modified quantization proposal by
Snyder \cite{Snyder}.

\subsection{Consequences of the quantization rules listed in Eq. (\ref{eq-18})}

There are two possibilities, a trivial and a non-trivial one:

\vskip 0.5cm
\noindent
(1) The trivial one recovers the classical quantization rule, 
when $y^\mu$ and thus also $p^y_\nu$ are set to zero.

\vskip 0.5cm
\noindent
(2) A non-trivial solution arises when $y^\mu$ and $p^y_\nu$ are not 
set to zero, which will be investigated in what follows. 

\vskip 0.5cm
The commutator
$\left[x^\mu , x^\nu \right]$ is equal to the negative 
of $\left(\frac{l}{c}\right)^2\left[y^\mu , y^\nu \right]$,
i.e., proportional to the minimal length element squared. 
Thus, a permitted ansatz is

\beqa
\left[ x^\mu , x^\nu \right] & = & \left( \frac{l}{c}\right)^2 {\bd C}^{\mu\nu}
~~~,
\label{eq-19}
\eeqa
where the symmetry property of the commutator on the left hand side
determines those of 
${\bd C}^{\mu\nu}$, which has to be an antisymmetric operator

\beqa
{\bd C}^{\mu\nu} & = & - {\bd C}^{\nu\mu}
~~~.
\label{eq-20}
\eeqa
Note, that ${\bd C}^{\mu\nu}$ is not specified yet. It may be a matrix with constant entries,
as used in the Moyal product \cite{obregon1,obregon2}, or it may be more general, being a function on
operators, as used in \cite{Snyder}.

(\ref{eq-19}) and (\ref{eq-20}) imply for the
commutator of the $y^\mu$

\beqa
\left[ y^\mu , y^\nu \right] & = & -{\bd C}^{\mu\nu} ~=~ {\bd C}^{\nu\mu}
~~~.
\label{eq-21}
\eeqa

For the commutators of the momentum operators, equal arguments apply:
Requiring the symmetry between coordinates and momenta under the
canonical transformation, which interchanges the coordinates and
momenta, leads to

\beqa
\left[ p^x_\mu , p^x_\nu \right] & = & \left( \frac{l}{c}\right)^2 {\widetilde  {\bd C}}_{\mu\nu}
\label{eq-21-1}
\eeqa
and for the pseudo-imaginary components

\beqa
\left[ p^y_\mu , p^y_\nu \right] & = & -{\widetilde  {\bd C}}_{\mu\nu} ~=~ {\widetilde  {\bd C}}_{\nu\mu}
~~~.
\label{eq-21-2}
\eeqa
In general, the ${\widetilde  {\bf C}}_{\mu\nu}$ are different to ${\widetilde  {\bd C}}^{\mu\nu}$ but we set them 
to ${\bf C}_{\mu\nu}$, which is the case when the ${\bf C}^{\mu\nu}$ are c-numbers.
This choice is in line with the reciprocity proposed by Born \cite{born1,born2}.

The second equation in (\ref{eq-18}) implies

\beqa
\left[ x^\mu , y^\nu \right] & = & {\bf B}^{\mu\nu}
~=~
\left[ x^\nu , y^\mu \right] ~ = ~ {\bf B}^{\nu\mu}
\nonumber \\
{\rm thus} &&
\nonumber \\
{\bd B}^{\mu\nu} & = & + {\bd B}^{\nu\mu}
~~~.
\label{eq-25}
\eeqa
The right hand side is either zero or proportional to a symmetric operator
${\bd B}^{\mu\nu}$, whose components also may be constant entries.

Similar for the momentum operators:

\beqa
\left[ p^x_\mu , p^y_\nu \right] & = & {\widetilde  {\bf B}}_{\mu\nu}
~=~
\left[ p^x_\nu , p^y_\mu \right] ~ = ~ {\widetilde  {\bf B}}_{\nu\mu}
\nonumber \\
{\rm thus} &&
\nonumber \\
{\widetilde  {\bd B}}_{\mu\nu} & = & + {\widetilde  {\bd B}}_{\nu\mu}
~~~,
\label{eq-25-b}
\eeqa
where as a special case one can set ${\widetilde  {\bf B}}_{\mu\nu}={\bf B}_{\mu\nu}$, which is valid if they are c-numbers and is consistent with the reciprocity between the
coordinates and momenta.

We repeat here the fifth equation in (\ref{eq-18})

\beqa
\left[ x^\mu , p^x_\nu \right] & = & i\hbar \delta_{\mu\nu} - 
\left(\frac{l}{c}\right)^2\left[ y^\mu , p^y_\nu \right]
~~~.
\label{eq-26}
\eeqa
Thus, the correction on the right hand side has to be of second order in
$\left(\frac{l}{c}\right)$. 

As mentioned above,
the detailed expressions of this correction and
the operators ${\bd C}^{\mu\nu}$ and ${\bd B}^{\mu\nu}$ are not fixed yet.
Also, the commutator $\left[y^\mu , p^y_\nu \right]$ in the last equation
has to be specified. This allows a great liberty for proposing several solutions.

\subsection{Comparison to H. S. Snyder's work}

In \cite{Snyder} the coordinates $x$, $y$ and $z$ are hermitian operators, whose spectrum is
invariant under the Lorentz transformation to other coordinates $x^\prime$, $y^\prime$
and $z^\prime$. It is noted that one solution corresponds to commuting coordinate 
operators with a continuous spectrum, which leads to the standard 
Quantum Mechanics. However,
another solution exists, where the coordinates do not commute. Snyder then constructs,
with the help of five variable $\eta_i$ (i=0,1,2,3,4), a representation of the coordinates

\beqa
x & = & il\left( \eta_4\frac{\partial}{\partial \eta_1} - \eta_1\frac{\partial}{\partial \eta_4} \right)
\nonumber \\
y & = & il\left( \eta_4\frac{\partial}{\partial \eta_2} - \eta_2\frac{\partial}{\partial \eta_4} 
\right)
\nonumber \\
z & = & il\left( \eta_4\frac{\partial}{\partial \eta_3} - \eta_3\frac{\partial}{\partial \eta_4} \right)
\nonumber \\
~~~ct & = & il\left( \eta_4\frac{\partial}{\partial \eta_0} + \eta_0\frac{\partial}{\partial \eta_4}
\right)
~~~.
\label{snyder1}
\eeqa
Thus, he extends the consideration to a 5-dimensional space, whose coordinates are
$\eta_\mu$, with $\mu$ = 0,1,2,3,4.

The coordinates have the typical structure of generators of a non-compact orthogonal group,
multiplied by $l$. Taking any component, its eigenvalue is $lm$, i.e., 
a multiple integer of the minimal length parameter $l$,
in the same way as the $L_z$ operator of the orbital angular momentum has integer eigenvalues.
The linear momentum operators are ratios of the $\eta_\mu$
(e.g., $p_x=\frac{\hbar}{l}\frac{\eta_1}{\eta_4}$) and, thus, commute as in standard Quantum Mechanics. 
{\it Note, that this introduces an asymmetry between the coordinates and momenta, while in
our approach the symmetry is maintained.}

Let us resume the quantization for the coordinates, 
as given in \cite{Snyder}, adapted to our notation
(the same signature for the metric was used)

\beqa
\left[x_i , x_k \right] & = & i \left(\frac{l^2}{\hbar}\right) L_{ik}
\nonumber \\
\left[x_0 , x_k \right] & = & i \left(\frac{l^2}{\hbar}\right) M_{0k}
~~~,
\label{S1}
\eeqa
with $i,k$=1,2,3 and where in our case $x_0 = ct$. The $L_{ik}$ are the angular momentum operators, while
$M_{0k}$ are the Lorentz boosts. 

The difference to our approach is that in H. S. Snyder's proposal
the Lorentz generators still fulfill the
commutation relations of the Lorentz group, while in our case this is not true any more,
when restricting to the x-space. We are convinced, ours is a justified and consistent
approach,
because due to the minimal length the Lorentz generators,
as a function in $x^\mu$ and $p^x_\nu$ only,
should not fulfill any more in the x-subspace 
the commutation relations of the Lorentz group, i.e., 
{\it Lorentz symmetry is apparently broken}. 

This also implies, that the commutators of the
coordinates with the Lorentz operators (and the same for the momentum operators)
are not any more the same as for the case when the minimal length is set to zero. 

In \cite{Snyder} the commutation relations of the coordinate with the momentum
operators are given by

\beqa
\left[ x_i , p_i \right] & = &
i\hbar \left[ 1+ \left( \frac{l}{\hbar}\right)^2 \left(p_i\right)^2 \right]
\nonumber \\
\left[ x_0 , p_0 \right] & = &
i\hbar \left[ 1- \left( \frac{l}{\hbar}\right)^2 \left(p_0\right)^2 \right]
\nonumber \\
\left[ x_i , p_0 \right] & = & \left[ x_0 , p_i \right] ~=~
i\hbar \left( \frac{l}{\hbar}\right)^2 p_i p_0
~~~.
\label{S2}
\eeqa

This looks very similar to our proposed quantization rules. The corrections
to the standard quantization are also proportional to $\left( \frac{l}{c}\right)^2$. 

To resume, the quantization given in \cite{Snyder} and as used in more recent
investigations \cite{obregon1,obregon2} 
are very similar to ours. However, there are
differences. The main common features and differences are

\begin{itemize}

\item The Lorentz symmetry is maintained in the 8-dimensional space but violated in the 4-dimensional x-space. 
The Lorentz operators, restricted to the dependence on $x^\mu$ and $p^x_\nu$, 
fulfill the commutation relations of the
Lorentz group only for a zero minimal length.
Contrary to that, in \cite{Snyder} the Lorentz symmetry is maintained
in the x-space.

\item In \cite{Snyder} the minimal length is introduced via the eigenvalues of a coordinate
operator,
which are kept invariant under Lorentz transformations.
In our contribution, the minimal length is introduced as a parameter, appearing in the 
pseudo-imaginary component of the coordinates and thus are also not affected by a Lorentz transformation.The coordinates are not discrete but continuous.

\item In our proposal the symmetry between the coordinates and momenta are maintained,
while in \cite{Snyder} it is broken.

\item The corrections to the commutation relations are very similar to those in \cite{Snyder}.
In \cite{Snyder} the expressions on the right hand side of the commutation relations of the
coordinates with coordinates, momenta with momenta and coordinates with momenta are well
defined, while in our case we still have liberty to choose them. 

\end{itemize}

\section{Pseudo-complex Quantum Mechanics in one dimension}

The intention of this chapter is to see how a pseudo-complex Quantum Mechanics might
be formulated, though the consideration will not be complete due to open questions.
Another point is to see, if there is a possibility to
measure the minimal length. For that purpose,
we consider two plane waves in one dimension, whose phase difference
is exactly $\pi$, i.e., in the classical theory these waves cancel exactly.
And this is the point: Look for a situation were the classical theory does give
zero contribution but in the pc-theory it is {\it different from zero}.
This example is easily extended to, e.g., a scattering of a plane wave by a crystal,
producing dark fringes in certain scattering angles.

One important observation is that in each zero-divisor component one defines a standard Quantum Mechanics, which we will investigate for the case of ´free particle motion, i.e.,
plane waves: The energy is given by the eigenvalue of the corresponding Hamiltonian, namely

\beqa
{\bd H} & = & \frac{{\bd P}^2}{2m} ~=~ \frac{\hbar^2 {\bd K}^2}{2m}
\nonumber \\
& \rightarrow &
\nonumber \\
E & = & \frac{\hbar^2 K^2}{2m}
~=~ \frac{\hbar^2 K_+^2}{2m}\sigma_+ + \frac{\hbar^2 K_-^2}{2m}\sigma_-
~~~,
\label{H1}
\eeqa
with the real mass $m$. In principle, it is interesting to assume the mass also as pseudo-complex, which however leads to the problem of how to interpret the pseudo-real and pseudo-imaginary components. For simplicity, we avoid this problem for the moment. The ${\bd K}$ is a pseudo-complex wave
number operator, with

\beqa
K & = & \frac{2\pi}{\Lambda}
~~~.
\label{H2}
\eeqa
The $\Lambda$ is a pc-wave length, which has units of a coordinate, 
where the pseudo-real component is taken as the normal
wavelength $\lambda$ and the pseudo-imaginary 
component is proportional to the {\it velocity} of it, i.e.,
the derivative of the wavelength of $\lambda$ with respect to a length element $ds$, which
can be the eigentime. Resumed, we have

\beqa
\Lambda & = & \lambda + \left(\frac{l}{c}\right)I\dot{\lambda}
~~~.
\label{H3}
\eeqa

With this, (\ref{H2}) acquires the form

\beqa
K & = & 2\pi \frac{\Lambda^*}{\Lambda \Lambda^*} ~=~ \frac{2\pi}{\lambda^2 - \left(
\frac{l}{c}\right)^2 \dot{\lambda}^2}
 \left[ \lambda - I \left( \frac{l}{c}\right) \dot{\lambda}\right]
\nonumber \\
& = & k_R + I\left(\frac{l}{c}\right) k_I
~~~,
\label{H4}
\eeqa
with

\beqa
k_R & = & \frac{2\pi}{\lambda} \frac{1}{\left[ 1-\left(\frac{l}{c}\right)^2 
\left( \frac{\dot{\lambda}}{\lambda}\right)^2\right]} ~=~ 
\frac{k}{\left[ 1-\left(\frac{l}{c}\right)^2 
\left( \frac{\dot{\lambda}}{\lambda}\right)^2\right]}
\nonumber \\
k_I & = & -\frac{k}{{\left[ 1-\left(\frac{l}{c}\right)^2 
\left( \frac{\dot{\lambda}}{\lambda}\right)^2\right]}} \frac{{\dot \lambda}}{\lambda}
~~~.
\label{H5}
\eeqa

Using $\dot{\lambda}$ = $\frac{d}{ds}\left(\frac{c}{\nu}\right)$ = $-\frac{c}{\nu^2}\dot{\nu}$ 
for a light wave, with $\nu$ as the frequency, we can rewrite $k_R$ and $k_I$ as

\beqa
k_R & = & \frac{k}{\left[ 1-\left(\frac{l}{c}\right)^2 
\left( \frac{\dot{\nu}}{\nu}\right)^2\right]}
\nonumber \\
k_I & = & \frac{k}{{\left[ 1-\left(\frac{l}{c}\right)^2 
\left( \frac{\dot{\nu}}{\nu}\right)^2\right]}} \frac{{\dot \nu}}{\nu}
\label{H6}
\eeqa

The interpretation of $\dot{\nu}$ is still pending and how to realize it, e.g., with
tunable lasers or materials which change their optical property as a function of the position,
thus, changing $\lambda$ and $\nu$ in space and/or time.

Finally, the energy is given by

\beqa
E & = & E_R + I E_I
\nonumber \\
E_R & = & \frac{\hbar^2}{2m} \left[ k_R^2 + \left( \frac{l}{c}\right)^2 k_I^2 \right]
\nonumber \\
E_I & = & \frac{\hbar^2}{m}\left( \frac{l}{c} \right) k_Rk_I
~~~.
\label{H7}
\eeqa

Analogous to classical Electrodynamics, one may adopt the standpoint that the physical
result is given by the real component of a complex expression. However, we will permit a more
general scenario, which leads to the problem on how to interpret the pseudo-imaginary component. 
We hope for a contribution from the readers.

For $k_I=0$ ($\dot{\nu}=0$) the real component in (\ref{H7}) 
of the energy reduces to the well known classical
result and the imaginary component $E_I=0$. Thus, the only differences appear when
there is a change in the frequency, which may be a time dependence or a change in terms
of a length parameter of a curve.

When only up to linear corrections in $\left(\frac{l}{c}\right)$ are considered, 
the pseudo-real and pseudo-imaginary components of the energy acquire a particular
simple form, namely

\beqa
E_R & \approx & \frac{\hbar^2 k^2}{2m}
\nonumber \\
E_I & \approx & \left(\frac{l}{c}\right)\frac{\hbar^2k^2}{m} \left( \frac{\dot{\nu}}{\nu}\right)
~~~.
\label{H8}
\eeqa

\subsection{Interpretation of probability distributions}

One question arises, namely
what is the meaning of the wave function in terms of probability? 
Here, we encounter three possibilities: 

\noindent
i) Define the probability distribution as

\beqa
\mid \Psi\mid^2 & = & \mid \Psi_+\mid^2 \sigma_+ + \mid \Psi_-\mid^2 \sigma_-
~~~,
\label{H9}
\eeqa
which is the complex norm in each pc-component, and the physical probability distribution
as

\beqa
\mid\mid \Psi \mid\mid^2 & = & \mid \Psi \mid_R^2 ~=~ 
\frac{1}{2} \left( \mid \Psi_+\mid^2 + \mid \Psi_-\mid^2 \right)
~~~.
\label{H10}
\eeqa
This definition also raises the question, of what is the meaning of the pseudo-imaginary
component

\beqa
\mid \Psi \mid_I^2 & = & 
\frac{1}{2} \left( \mid \Psi_+\mid^2 - \mid \Psi_-\mid^2 \right)
\label{H11}
\eeqa
of the pc-wave function?

\noindent 
ii) Use the {\it euclidean Norm}

\beqa
\mid\mid \Psi \mid\mid_E & = & \sqrt{ \mid \Psi_R\mid^2 + \mid \Psi_I\mid^2}
\label{euk}
\eeqa
with

\beqa
\Psi_R & = & \frac{1}{2} \left( \Psi_+ + \Psi_- \right)
\nonumber \\
\Psi_I & = & \frac{1}{2} \left( \Psi_+ - \Psi_- \right)
~~~.
\label{euk-1}
\eeqa

Using (\ref{euk-1}) in (\ref{euk}) leads to

\beqa
\mid\mid \Psi \mid\mid_E & = & \sqrt{ 
\frac{1}{2}\left( \mid \Psi_+\mid^2 + \mid \Psi_-\mid^2 \right)
}
~~~,
\label{H11-b}
\eeqa
which is the same as in (\ref{H10}), but does not require the interpretation
of $\mid \Psi \mid_I$! 
The definition of the euclidean norm seems to be consistent with i).

\noindent
iii) Or, define the probability as multiplying the pc-probability by its pc-conjugate, namely

\beqa
\mid\mid \Psi \mid\mid^4 & = & \left( \mid \Psi_+\mid^2 \sigma_+ + 
\mid \Psi_-\mid^2 \sigma_- \right)
\left( \mid \Psi_+\mid^2 \sigma_+ + 
\mid \Psi_-\mid^2 \sigma_- \right)^*
\nonumber \\
& = & \left( \mid \Psi_+\mid^2 \sigma_+ + 
\mid \Psi_-\mid^2 \sigma_- \right)
\left( \mid \Psi_+\mid^2 \sigma_- + 
\mid \Psi_-\mid^2 \sigma_+ \right)
\nonumber \\
& = & \mid \Psi_+\mid^2 \mid \Psi_-\mid^2
~~~.
\label{H12}
\eeqa
The probability is then the square root of this, i.e.

\beqa
\mid\mid \Psi\mid\mid^2 & = & \mid \Psi_+\mid \mid \Psi_-\mid
~~~
\label{H13}
\eeqa

Which of the three definitions makes sense is not yet decided and we hope a constructive
contribution from the reader. {\it I tend to use the euclidean norm}.

\subsection{A simple example}

Let us assume that the wave function is a plane wave, with pseudo-complex argument, that is

\beqa
{\rm cos}({\bd K}\cdot {\bd X})
~~~,
\label{eq-27}
\eeqa
where ${\bd K}$ is a pc-vector for the wave number and ${\bd X}$ is a pc-coordinate vector,
i.e., in both cases just the spatial part, equal to
$K_iX^i$. This is for standing waves.

We use

\beqa
X & = & x + I\left(\frac{l}{c}\right) u ~=~ x+I\left(\frac{l}{c}\right) c
\nonumber \\
K & = & k + I \left(\frac{l}{c}\right)k_I
~~~.
\label{eq-28}
\eeqa
We denoted the pseudo-real components of the coordinate and the wave vector by $x$ 
and $k$ respectively. 
The pseudo-imaginary component $y$ is now interpreted (see section 3) as the velocity.
The "four" velocity component is set to the one of light,
assuming that this is the velocity of propagation of the wave. The pseudo-imaginary component
$k_I$ of $K$ is interpreted as given in (\ref{H6}).

Rewriting it in terms of the pseudo-real and pseudo-imaginary component, we obtain

\beqa
&
\Psi ~=~{\rm cos} \left( \left[kx+\left(\frac{l}{c}\right)^2 ck_I\right]+\left(\frac{l}{c}\right)
I\left[x k_I + ck\right]\right)
&
\nonumber \\
& = &
\nonumber \\
&
{\rm cos} \left( \left[kx+\left(\frac{l}{c}\right)^2 ck_I\right] \right)
{\rm cos}\left( \left(\frac{l}{c}\right)\left[x k_I + ck\right]\right)
&
\nonumber \\
&
-I
{\rm sin} \left( \left[kx+\left(\frac{l}{c}\right)^2 ck_I\right] \right)
{\rm sin}\left( \left(\frac{l}{c}\right)\left[x k_I + ck\right]\right)
&
\nonumber \\
&
~=~ \Psi_R + I \Psi_I
~~~.
&
\label{eq-29}
\eeqa
We have used that the ${\rm cos}$-function has only even powers of the argument,
while in the ${\rm sin}$-function there are only odd powers. Taking into account that
$I^{2n}=1$ and $I^{2n+1}=I$, finally leads to the last expression.
From (\ref{eq-29}) one infers that the pseudo-imaginary component is already of the order of 
$\left( \frac{l}{c} \right)$, due to the second factor 
${\rm sin} \left(\left(\frac{l}{c}\right)
\left[xk_I+ck\right]\right)$ $\sim$ $\left(\frac{l}{c}\right)$.

To this, a wave of the same form, but shifted by $x$ $\rightarrow$ $x+\frac{\pi}{k}$
is added. A direct calculation shows that the 
the leading contribution of the real part is proportional to 
$\left(\frac{l}{c}\right)^2$, while the pseudo-imaginary part stays proportional to
$\left(\frac{l}{c}\right)$. We obtain up to linear order in $\left(\frac{l}{c}\right)$

\beqa
\Psi_R & \approx & 0
\nonumber \\
\Psi_I & \approx & \left( \frac{l}{c}\right) \left( \frac{k_I}{k}\pi\right)
{\rm sin} \left( kx \right)
~~~.
\label{eq-29-a}
\eeqa

Thus, the euclidean norm square gives

\beqa
\mid\mid \Psi \mid\mid^2 ~=~ \mid\mid \Psi \mid\mid_E^2 & \approx & 
\left( \frac{l}{c}\right)^2 \left( \frac{k_I}{k}\pi\right)^2
{\rm sin}^2 \left( kx \right)
\nonumber \\
& = &
\left( \frac{l}{c}\right)^2 \left( 
\pi
\frac{{\dot \nu}}{\nu}
\right)^2
{\rm sin}^2 \left( kx \right) 
~~~,
\label{eq-29-b}
\eeqa
where the $k_I$ of (\ref{H6}) was substituted and only up to first order terms
in $\left(\frac{l}{c}\right)$ are taken into account.

This result shows that when classically a wave is annihilated by a negative interference,
in the pc-theory a remnant wave is maintained, i.e., there is no complete annihilation.
This can be used to measure the minimal length, when $\nu$, $\dot{\nu}$ are known.

However, the practical problem lies in the numerical values:
The correction is proportional to $\left( \frac{l}{c} \right)^2$, which has probably an extremely
small value. Even for $l=10^{-16}$cm, the current limit in LHC experiments at CERN,
this factor is at least $10^{-53}$sec$^2$, not having included the value of $k_I$
$\sim$ $\frac{\dot{\nu}}{\nu}$ and also not having multiplied by the intensity of the wave,
which improves the situation a bit.
It is expected that this non-zero effect is possible not observable with the
current technology available.

\section{Conclusions}

I have shown that extending algebraically the 4-dimensional flat space
coordinates to pseudo-complex coordinates, i.e., to an 8-dimensional space, continuous
symmetries, like the Lorentz symmetry, are maintained, but still one can introduce
a minimal length scale.

Although, in this 8-dimensional space the quantization rules are equal to the classical one,
in the 4-dimensional subspace the commutation rules of the coordinates are those one expects when a minimal length is introduced. Thus, the complicated commutation relations have a trivial
picture in the 8-dimensional space. This may provide a simplified picture to quantize space
with a minimal length present, using the pc-formulation, and hopefully also indicates on how gravity can be quantized.

We compared the new quantization proposal to the one of H. S. Snyder \cite{Snyder} and showed that
the mathematics is similar, though at certain important points, differences appear. For example, in \cite{Snyder} the generators of the Lorentz group satisfy the known commutation
relations, while in our case there is a correction of the order
of the minimal length squared. Thus, the Lorentz symmetry is {\it effectively broken}
in the x-space. Note, that
this is only apparent, because in the 8-dimensional space this symmetry is still present.
As a working rule, one should perform all calculations in the 8-dimensional space and at the end
deduce the real observables.

In the last section, we realized the first steps toward a 1-dimensional pc-Quantum Mechanics.
Apart from known results, new structures arise which still defy a proper interpretation.
We hope, that this contribution encourages some of the readers to seek for it.

As an example, the overlap of two plane waves was considered which are out of phase by
$\pi$. Though, in the classical picture these waves cancel each other, due to the presence of
a minimal length the amplitude remains different from zero. Unfortunately, the non-zero
contribution of the amplitude squared is too small that there is some hope to see it
with the technology presently available.

I am aware that a lot has to be done yet. This contribution serves mainly to start a conversation
toward a new quantization proposal, with important implications.

\section*{Acknowledgment}
The financial help from DGAPA (PAPIIT No. IN103212) and CONACyt are acknowledged. P.O.H. also
wants to thank the {\it Frankfurt Institute for Advanced Studies} (J.W. von-Goethe
Universit\"at, Frankfurt am Main, Germany) for its hospitality, where part of the present
study was realized. especially discussions with Prof. Dr. W. Greiner and Dr. M. Sch\"afer
are acknowledged.

\section*{APPENDIX A: Structure of the generators of the pc-Lorentz group}

The $\sigma_\pm$ components of the generators of the Lorentz group are given by

\beqa
L_{\mu\nu} & = & L^+_{\mu\nu}\sigma_+ + L^-_{\mu\nu}\sigma_-
\nonumber \\
L^\pm_{\mu\nu} & = & X^\pm_\mu P^\pm_\nu - X^\pm_\nu P^\pm_\mu
~~~.
\label{Lo1}
\eeqa
Using $X^\pm_\mu = x_\mu \pm \left(\frac{l}{c}\right)y_\mu$
and $P^\pm_\mu = p^x_\mu \pm \left(\frac{l}{c}\right) p^y_\mu$,
we get for the real and pseudo-imaginary part of the pc-generator

\beqa
L^R_{\mu\nu} & = & \frac{1}{2} \left( L^+_{\mu\nu} + L^-_{\mu\nu} \right)
\nonumber \\
& = &
\left( x_\mu p^x_\nu -x_\nu p^x_\mu\right) 
+ \left(\frac{l}{c}\right) \left( y_\mu p^y_\nu - y_\nu p^y_\mu\right)
\nonumber \\
L^I_{\mu\nu} & = & \frac{1}{2} \left( L^+_{\mu\nu} - L^-_{\mu\nu} \right)
\nonumber \\
& = &
\left(\frac{l}{c}\right) \left( x_\mu p^y_\nu - x_\nu p^y_\mu 
+ y_\mu p^x_\nu- y_\nu p^x_\mu\right)
~~~.
\label{Lo2}
\eeqa
As can be seen, the pseudo-real part $L^R_{\mu\nu}$
resembles most to a Lorentz group generator,
with the additional similar contribution of operators defined in the $y$ coordinates and
momenta of the order $\left(\frac{l}{c}\right)$. 
Calculating the commutators of $L^R_{\mu\nu}$ is rather cumbersome. 

Here, we will only determine the commutation relations of the

\beqa
{\cal L}_{\lambda\delta} & = & x_\lambda p^x_\delta - x_\delta p^x_\lambda
~~~.
\label{Lo3}
\eeqa

First, using the quantization rules (\ref{eq-18}), we determine

\beqa
\left[ x_\mu , {\cal L}_{\lambda\delta}\right] & = &
i\hbar \left[ \eta_{\mu\delta} x_\lambda - \eta_{\mu\lambda} x_\delta\right]
\nonumber \\
&& + \left(\frac{l}{c}\right)^2
\left\{
x_\delta \left[y_\mu , p^y_\lambda\right] - x_\lambda\left[y_\mu , p^y_\delta\right]
+ \left[y_\mu , y_\delta\right] p^x_\lambda - \left[ y_\mu , y_\lambda\right] p^x_\delta
\right\}
\nonumber \\
\left[ p^x_\nu , {\cal L}_{\lambda\delta}\right] & = &
-i\hbar \left[ \eta_{\lambda\nu} p^x_\delta - \eta_{\nu\delta} p^x_\lambda\right]
\nonumber \\
&& + \left(\frac{l}{c}\right)^2
\left\{
-x_\lambda \left[p^y_\nu , p^y_\delta\right] + x_\delta\left[p^y_\nu , p^y_\lambda\right]
+ \left[y_\lambda , p^x_\nu\right] p^x_\delta - \left[ y_\delta , p^x_\nu\right] p^x_\lambda
\right\}
~~~,
\nonumber \\
\label{Lo4}
\eeqa
where we have also made use of the relations given in (\ref{eq-18}).

From this, we obtain the commutator of $x_\mu p^x_\nu$ with ${\cal L}_{\lambda\xi}$

\beqa
\left[ x_\mu p^x_\nu , {\cal L}_{\lambda\delta}\right] & = &
x_\mu \left[ p^x_\nu , {\cal L}_{\lambda\delta}\right] + \left[ x_\mu ,
{\cal L}_{\lambda\delta} \right] p^x_\nu
~~~.
\label{Lo5}
\eeqa

Using the result of the former equation and subtracting to it the same expression
with $\mu$ and $\nu$ interchanged, we obtain finally the following structure

\beqa
\left[x_\mu p^x_\nu - x_\nu p^x_\mu , {\cal L}_{\lambda\delta}\right] ~=~
\left[ {\cal L}_{\mu\nu} , {\cal L}_{\lambda\delta}\right] & = &
i\hbar \left(
\eta_{\lambda\nu} {\cal L}_{\delta\mu}
+\eta_{\delta\nu} {\cal L}_{\mu\lambda}
+\eta_{\mu\lambda} {\cal L}_{\nu\delta}
+\eta_{\delta\mu} {\cal L}_{\lambda\nu}
\right)
\nonumber \\
&&
+\left(\frac{l}{c}\right)^2 {\bd F}_{\mu\nu\lambda\delta}\left( x, p^x, y, p^y \right)
~~~,
\label{Lo6}
\eeqa
Where ${\bd F}$ can be deduced, using
(\ref{Lo4}) and (\ref{Lo5}).
The first term, with the factor $i\hbar$ is the one which is expected for the commutator
of Lorentz generators. However, there is a correction due to the minimal length and it
is proportional to the minimal length squared, i.e., a real small correction.

Thus, the operators ${\cal L}_{\mu\nu}$ can be considered as approximate Lorentz
generators and only for large acceleration,
which is proportional to the inverse of the minimal length
\cite{caianiello}, the correction to the commutator
becomes appreciable and the Lorentz symmetry is apparently violated on the level 
of the $x$-space.

\end{document}